# OPTIMISING PHOTOTHERMAL SILVER NANOPARTICLES FOR EFFICIENT LIGHT-ACTIVATED SHAPE MEMORY RESPONSE IN AgNP-POLYMER COMPOSITES

K. Krizmane, M. Dile, E. Einbergs, V. Vitola*, A. Knoks,
B. Hamawandi, A. Zolotarjovs

Institute of Solid State Physics, University of Latvia,
8 Ķengaraga Str., Riga, LV-1063, LATVIA
*e-mail: virginija.vitola@cfi.lu.lv

The study presents the synthesis and integration of silver nanoparticles (AgNPs) into a polyurethane (PU) matrix to create a light-activated shape memory polymer (SMP). AgNPs were synthesized using a microwave-assisted solvothermal method with polyvinylpyrrolidone (PVP) as a stabilizer, where varying the molecular weight of PVP and the $AgNO_3$/PVP ratio influenced nanoparticle size and distribution. Characterisation via scanning electron microscopy (SEM) and dynamic light scattering (DLS) confirmed the irregular morphology and size consistency of AgNPs, with smaller particles exhibiting narrow size distributions and enhanced photothermal response. Among the samples, PVP with an average molecular weight ($M_n$) of 10,000 and an $AgNO_3$/PVP mole ratio of 14.7 demonstrated optimal UV-blue light absorption, which facilitated efficient local heating under irradiation. The AgNP-PU composite exhibited a reliable shape memory effect when exposed to UV-blue light.

**Keywords:** *Microwave-assisted synthesis, photothermal effect, polymer composite, shape memory, silver nanoparticles.*

# 1. INTRODUCTION

In recent years, SMPs have increasingly become pivotal in the field of actuating applications, ranging from everyday products to biomedical [1]–[3] and aeronautical devices [4], [5]. An important turning point in the usability of these materials has been the possibility to 3D-print the desired final shape. Four-dimensional (4D) printing, an advanced evolution of 3D printing, has acquired substantial interest across various research domains, including smart materials and biomedical research. This technique enables a 3D printed structure to transform its shape over time in response to specific stimuli such as light, heat, electrical current, magnetic fields, or osmotic pressure [6]–[9]. SMPs are types of polymers that can maintain a temporary shape due to their network points and flexible chains. The shape memory effect in thermal SMPs is induced by subjecting the material to a thermal stimulus that triggers a transition between its temporary and initial shape states. When heated to a specific temperature, above the polymer's glass transition temperature ($T_g$), the polymer chains gain sufficient mobility, allowing the material to revert to its initial shape (Fig. *1*). This thermally activated process is governed by the polymer's molecular structure, which can be engineered to exhibit reversible phase transitions. Upon cooling below the glass transition temperature, the polymer retains the recovered shape, effectively "locking" in the morphology due to restricted chain mobility [10].

By optimising the thermal and mechanical properties of the polymer, it is possible to achieve controlled and repeatable shape memory behaviour suitable for applications in biomedical devices [1], [11], [12], actuators [13], [14], and smart materials [15], [16]. There are different thermal SMPs, i.e., PU, which is valued for its versatility, biocompatibility, and potential biodegradability when modified with eco-friendly components. PU's biocompatibility makes it particularly suitable for medical applications, including flexible foams and durable elastomers that come into direct contact with biological tissues [17]–[19]. Polylactic acid (PLA), recognized for its robust mechanical strength, biodegradability, biocompatibility, and non-toxicity, has been increasingly used in biomedical applications in recent years [20], [21]. Poly-ε-caprolactone (PCL), an affordable semi-crystalline aliphatic polyester, is utilised in bone tissue engineering for its drug delivery and wound healing properties [22]–[24].

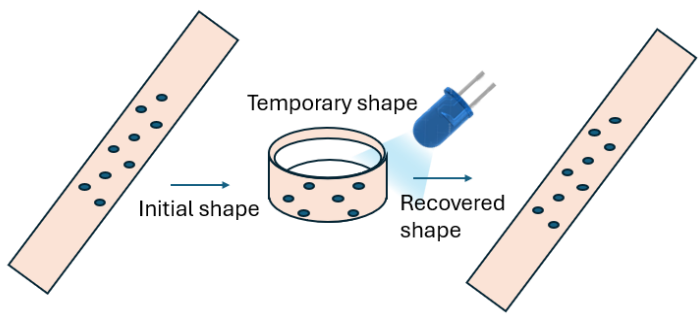

*Fig. 1.* A schematic drawing of shape memory composite in action.

Beyond traditional thermal activation, the shape memory effect in thermal SMPs can be achieved through the photothermal effect, where embedded photothermal nanoparticles enable light-activated heating (Fig. *1*). Specifically, by incorporating photothermal particles such as AgNPs into the polymer matrix, shape recovery can be triggered via light exposure [25]. This process leverages the nanoparticles' capacity to absorb light, particularly in the UV or visible spectrum, converting it into localised heat through surface plasmon resonance (SPR) [26], [27]. The heat generated by these nanoparticles can raise the polymer's temperature above its glass transition or melting temperature, initiating the shape memory effect without direct bulk heating [28]. In this study, we synthesized AgNPs using PVP as a stabilizing agent, exploring the influence of PVP molecular weight and the $AgNO_3$/PVP ratio on nanoparticle properties. The length of the PVP polymer chains, which correlates with its molecular weight, plays a crucial role in nanoparticle stabilization and size, directly impacting light absorption efficiency. Our findings demonstrated that AgNPs synthesized with PVP ($M_n$ = 10,000 g/mol) at an $AgNO_3$/PVP ratio of 14.7 exhibited the highest photothermal activity, achieving optimal UV-blue light absorption and efficient heat transfer to the PU matrix. This method allows for precise, spatially resolved activation of the shape memory effect via light, presenting promising applications in remote-controlled actuation suitable for biomedical applications.

## 2. EXPERIMENTAL

### 2.1. Materials

Silver nitrate ($AgNO_3$, purity ≥ 99 %, Lach-Ner) was used as an Ag precursor. Polyvinylpyrrolidone (PVP) with three different molecular weights ($M_n$ = 10,000, 29,000 and 90,000 g/mol, Sigma-Aldrich) was used as surface stabilizer for AgNPs. Ethylene glycol (EG, $C_2H_6O_2$, purity ≥ 99 %, Sigma Aldrich) was used as both a solvent and a reducing agent. Deionized water (DIW, $\rho = 18.2$ MΩ cm at 25 °C, total organic carbon up to 20 ppb, microorganisms < 10 CFU mL$^{-1}$, heavy metals < 0.01 ppm, silicates < 0.01 ppm, and total dissolved solids < 0.03 ppm) and acetone (purity ≥ 99.5%, Sigma-Aldrich) were used to precipitate the synthesized AgNPs, and ethanol ($C_2H_5OH$, purity 96%; Supelco) was used to redisperse the nanoparticles for later incorporation into the polymer matrix. To prepare the polymer matrix for AgNP embedding, PU-22 universal polyurethane polymer resin and HU-22 universal hardener (Uzlex, WMT Baltic) were utilised. All chemicals were of analytical grade and used as received, without further purification.

### 2.2. Synthesis of AgNPs

Ag nanoparticles stabilized by PVP were synthesized via a microwave-assisted polyol process, following the method developed by Lalegani et.al. [19]. To ensure efficient transfer of microwave radiation for rapid and uniform heating, each sample was prepared individually, with only one vial placed in the microwave synthesis system at a time.

First, a specified amount of PVP poly-

mer was dissolved in EG under constant stirring. For higher molecular weight PVP, heating was applied to promote dissolution. After cooling the solution to room temperature, a measured quantity of AgNO$_3$ was dissolved, and the mixture was transferred to a 70 mL PTFE container in a microwave synthesis system. The specific quantities of reagents used in this synthesis are listed in Table 1.

AgNP synthesis was performed in a Milestone synthWAVE microwave reactor. The system operates at 2.45 GHz frequency with variable power from 0 to 100 % (1.5 kW). The reaction was carried out at 160 °C for 10 min under constant stirring speed (60 % of maximum stirring speed) in an inert atmosphere (N$_2$ gas) and 40 bar pre-load pressure. The target temperature was reached within 2.5 minutes. After microwave processing, the reaction mixture was naturally cooled down to room temperature, and transparent colloidal solutions with colours ranging from yellow to brownish red was obtained. As the reducing ability of ethylene glycol (EG) becomes negligible at room temperature, effectively slowing down nanoparticle growth, the synthesized AgNPs were stored in EG for further characterisation. For subsequent dispersion of AgNPs in ethanol for embedding in polymer matrix, the samples were alternately washed several times with DIW and acetone.

**Table 1.** Amount of Reagents Used in AgNP Synthesis with the Ratio and Obtained Sample Colour

| Sample name | M$_n$ of PVP (g/mol) | PVP (g) | AgNO$_3$ (g) | EG (mL) | Mole ratio AgNO$_3$/PVP | Sample colour |
|---|---|---|---|---|---|---|
| Ag_1 | 10,000 | 0.10 | 0.1 | 40 | 58.8 | brownish red |
| Ag_2 | 10,000 | 0.40 | 0.1 | 40 | 14.7 | red |
| Ag_29 | 29,000 | 0.29 | 0.1 | 40 | 58.8 | brownish yellow |
| Ag_90 | 90,000 | 0.90 | 0.1 | 40 | 58.8 | yellow |

## 2.3. Fabrication of Polymer/Nanoparticle Composite

AgNP sample labelled Ag_2 was used to fabricate the PU/AgNP composite. 0.9 mL of 0.03 M AgNP solution in ethanol was added to a mixture of 1 g PU-22 and 1 g HU-22. The mixture was stirred with a magnetic stirrer for 5 min to ensure even distribution of nanoparticles throughout the polymer. The polymer mixture was then cast onto a polystyrene plate and left to dry at room temperature for 2 days. AgNP/PU composite containing 0.15 % AgNP was obtained. We tested the glass transition temperature of the employed polymer, and it was found to be 38 °C.

## 2.4. Methods of Characterisation

The morphology of the Ag NPs samples was characterised by scanning electron microscopy (SEM) (Tescan Lyra, Brno-Kohoutovice, Czech Republic), operated at 12 kV. A drop of the AgNP solution in ethylene glycol was placed onto a silicon wafer and air-dried before SEM analysis. Additional measurements were performed on the AgNPs in ethylene glycol solution. Particle size distribution was characterised by dynamic light scattering (DLS) using Litesizer 500 (Anton Paar). Optical density was measured with Cary 7000 Universal Measurement Spectrophotometer (Agilent, USA). The nanoparticle samples, at a concentration of 0.015 M, were diluted in EG in a 1:1 ratio. OD measurements were conducted using a quartz cuvette (spectral range 200–2,500 nm, pathlength 10 mm). Temperature changes under UV-blue light exposure were recorded using a Unisense x-5 UniAmp with TP-200 glass thermocouple, coupled with 420 nm LED.

## 3. RESULTS AND DISCUSSION

The $M_n$ of PVP polymer is directly connected to its chain length, where higher molecular weight compounds have longer polymer chains. In the synthesis process of silver nanoparticles, PVP is used to coat and stabilise nanoparticles. The chain length of PVP affects the radius of the resulting nanoparticles and can lead to variations in absorption. In this research, we experimented with different molecular weight PVP polymers to determine how the length of the attached polymer chain influenced the absorption of UV-blue light by silver nanoparticles. Additionally, the PVP to $AgNO_3$ ratio was varied to obtain the best conditions for synthesis (Table 1). Our goal was to achieve effective UV-blue light absorption due to the SPR of silver nanoparticles.

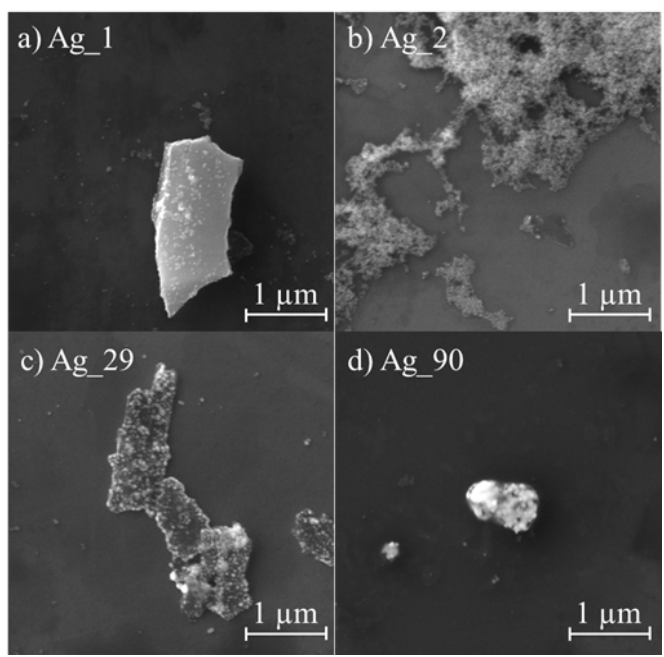

*Fig. 2.* SEM images of AgNP prepared with PVP; a) and b) $M_n$ = 10,000 g/mol, c) $M_n$ = 29,000 g/mol, d) $M_n$ = 90,000 g/mol.

The morphology of the prepared samples was determined using SEM analysis. Fig. 2. shows the formation of irregular Ag NP with sizes ranging between 20 nm and 60 nm. In samples labelled Ag_1, Ag_29, Ag_90, particles are capped with a surrounding layer. This layer most likely consists of organic compounds, more specifically, PVP polymer chains. SEM images show agglomerates in samples where higher $M_n$ PVP was used ($M_n$ = 29,000, 90,000 g/mol), suggesting these particles may also agglomerate in EG solution. In contrast, Ag_1 and Ag_2, prepared with lower $M_n$ PVP (10,000 g/mol), show less agglomeration in SEM images, particularly for sample Ag_2, where $AgNO_3$/PVP ratio was 14.7. This observation implies that the particles are likely well dispersed in EG solution without forming agglomerates.

**Table 2.** DLS Measurements of AgNPs Showing Hydrodynamic Diameter and Size Distribution (PDI)

| Sample name | Hydrodynamic diameter, nm | PDI, % |
|---|---|---|
| Ag_1 | 44 | 26 |
| Ag_2 | 41 | 0.18 |
| Ag_29 | 394 | 22 |
| Ag_90 | 1246 | 29 |

DLS measurements (Table 2) confirm these observations. For Ag_90, the measured hydrodynamic diameter is 1246 nm, likely representing organic-coated agglomerates rather than individual Ag NPs. Similarly, Ag_29 has a hydrodynamic diameter of 394 nm. In contrast, Ag_1 and Ag_2 show similar nanoparticle sizes (44 and 41 nm, respectively) but differ in polydispersity index (PDI), with Ag_2 exhibiting a narrower size distribution (PDI = 0.18 %). SEM images support this, showing less agglomeration in Ag_2 and suggesting that PVP provides better stabilization of nanoparticles in this sample.

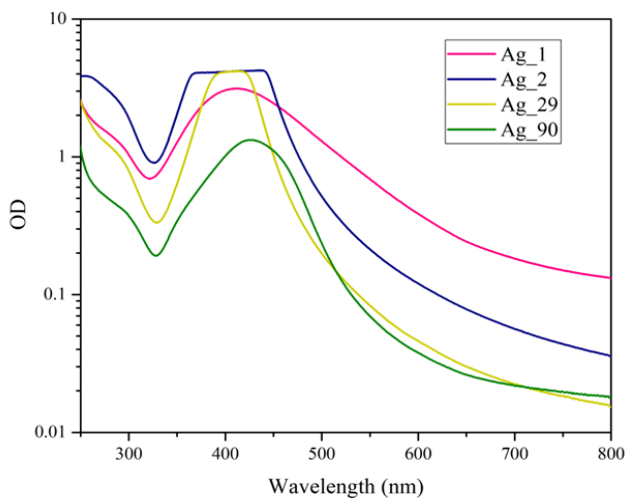

*Fig. 3.* The optical density (OD) of AgNP solution (1:1) in ethylene glycol.

The OD of the AgNP suspension in EG was measured using a spectrophotometer (Fig. 3.), employing EG in the reference beam to account for its baseline absorption. The Ag_2 sample ($M_n$ PVP = 10,000 g/mol, $AgNO_3$ to PVP mole ratio 14.7) exhibited the highest absorption in the UV and blue spectral range, with a prominent peak around 400–450 nm, which was characteristic due to their SPR. Typically, AgNPs with sizes around 10–50 nm display a strong absorption band in this region, as the collective oscillation of conduction electrons on their surface resonates with incident light, particularly in the visible and near-UV ranges. This SPR peak intensity and position are influenced by nanoparticle size, shape, and the refractive index of surrounding media. The observed high OD in the Ag_2 sample suggests efficient UV-blue light absorption, which is ideal for photothermal applications requiring localised heating and enhanced responsiveness in SMPs.

Following the optical density measurements, we assessed the photothermal heating effect of the AgNP solution in EG by coupling a temperature sensor with a blue light-emitting diode. The resulting temperature changes are depicted in Fig. 4.

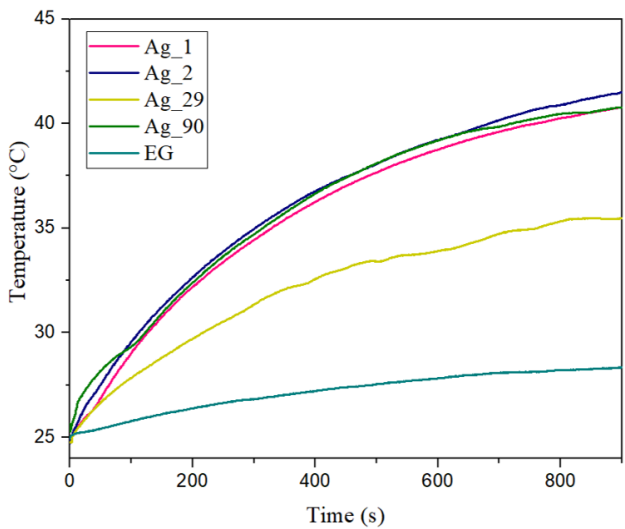

*Fig. 4.* Temperature change for different AgNP samples under blue light exposure.

This setup allowed us to monitor the temperature changes induced by light absorption directly within the AgNP solution. The temperature profiles of different AgNP samples in EG demonstrated a similar trend: upon blue light exposure, the temperature initially increased, with each sample exhibiting a distinct heating rate. However, after a certain period, the temperature reached a plateau, indicating that the heating rate due to light absorption balanced with the cooling rate to the surrounding environment. This equilibrium point reflects the thermal stability achievable with each sample under continuous illumination, determined by factors such as nanoparticle concentration, size, and light absorption efficiency. The variations in heating rates before reaching the plateau also highlight differences in the photothermal conversion efficiency of each sample. The Ag_2 and Ag_90 samples exhibited the fastest heat-ing rate and the largest plateau temperature, which is also above the transition temperature for the PU polymer – that means these two NP samples could be used for the light-activated shape memory polymer.

The nanoparticle sample Ag_2, which showed the best heating efficiency, was embedded in the polymer for the PU-nanoparticle composite, and the shape memory effect was tested and compared to a polymer without nanoparticles. Both samples were cut into equal size: thin, straight strips. They were submerged in 38 ºC hot water and formed in the temporary shape – a ring, and after that, the samples were rapidly cooled in cold water to fix the temporary shape. Both samples showed a stable temporary shape. After that, an LED (40 mW, 420 nm, Thorlabs) was placed 5 cm from the samples, and the shape recovery was observed. The polymer without nanoparticles did not recover its original shape. It remained in ring form, but the PU-AgNP composite unrolled to the original shape, thus demonstrating the shape memory effect (Fig. 5.).

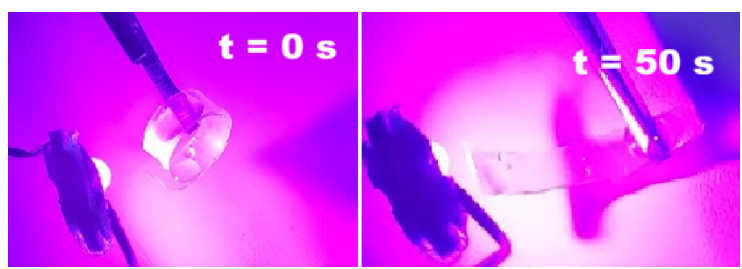

*Fig. 5.* Unfolding of the shape-memory PU/AgNP composite under 420 nm light stimulus; rolled polymer (t = 0s), unrolled polymer (t = 50s).

## 4. CONCLUSIONS

Silver nanoparticles stabilised by polyvinylpyrrolidone were successfully synthesized using a microwave-assisted solvothermal method. The molecular weight and concentration of PVP played a critical role in determining the size and distribution of the nanoparticles, with different $AgNO_3$/PVP ratios resulting in varied nanoparticle characteristics. Dynamic light scattering and scanning electron microscopy analyses confirmed the size and morphology of the AgNPs. SEM images revealed PVP coated irregularly shaped silver nanoparticles, while DLS measurements indicated that samples synthesised with lower molecular weight PVP had a narrower size distribution, suggesting that the molecular weight of PVP affects synthesized nanoparticle distribution.

Notably, the smallest size distribution was observed in the Ag_2 sample, where the $AgNO_3$/PVP ratio was 14.7, differing from other samples. This indicated that the PVP concentration significantly influences particle agglomeration, which is critical for achieving efficient light absorption and photothermal conversion. Among the synthesized samples, Ag_2 sample exhibited optimal photothermal activity. This sample demonstrated high UV-blue light absorp-tion efficiency, which is attributed to its surface plasmon resonance. This enhanced photothermal effect is crucial for light-activated shape-memory applications. When integrated into a polyurethane matrix, Ag_2 nanoparticles enabled the composite to respond to UV-blue light, thus triggering the shape memory effect. It was tested that while a PU sample without nanoparticles did not revert to its original shape, the PU-AgNP composite demonstrated successful shape recovery under UV-blue light, proving the effectiveness of AgNPs in inducing light-activated shape memory effect. This study supports that the potential of AgNP-integrated shape memory polymers for applications in remote-controlled actuation is advantageous. Future research should focus on optimising the concentration of AgNPs within the polymer to improve heating efficiency and further enhance the material's shape-memory capabilities.

## ACKNOWLEDGEMENTS

Virginija Vitola acknowledges the project LZP-2023/1-0521. The authors acknowledge the Institute of Solid State Physics of the University of Latvia, which, as a center of excellence, has received funding from the European Union framework program Horizon 2020 H2020-WIDESPREAD-01-2026-2017-TeamingPhase2 within grant agreement No. 739508 of the CAMART2 project.